%% file: main.tex
\documentclass[sigconf]{acmart}

\usepackage{graphicx}
\usepackage{multirow}
\usepackage{subfigure}
\usepackage{soul}
\usepackage{longtable}

\AtBeginDocument{%
  \providecommand\BibTeX{{%
    \normalfont B\kern-0.5em{\scshape i\kern-0.25em b}\kern-0.8em\TeX}}}

\setcopyright{rightsretained}
\copyrightyear{2021}
\acmYear{2021}
\acmDOI{10.1145/3463676.3485611}

\acmConference[WPES '21]{Proceedings of the 20th Workshop on Privacy
in the Electronic Society}{November 15, 2021}{Virtual Event, Republic of Korea}
\acmBooktitle{Proceedings of the 20th Workshop on Privacy in the
Electronic Society (WPES '21), November 15, 2021, Virtual Event, Republic of Korea}

\begin{document}

\title{Cookie Banners, What’s the Purpose?\\ Analyzing Cookie Banner Text Through a Legal Lens}

\author{Cristiana Santos*}
\affiliation{%
  \institution{Utrecht University}
  \country{Netherlands}
}
\author{Arianna Rossi*}
\affiliation{%
  \institution{SnT, University of Luxembourg}
  \country{Luxembourg}
}
\author{Lorena Sánchez Chamorro}
\affiliation{%
  \institution{University of Luxembourg}
  \country{Luxembourg}
}
\author{Kerstin Bongard-Blanchy}
\affiliation{%
  \institution{University of Luxembourg}
  \country{Luxembourg}
}
\author{Ruba Abu-Salma}
\affiliation{%
  \institution{King’s College London}
  \country{UK}
}

\renewcommand{\shortauthors}{Santos, et al.}

\begin{abstract}
A cookie banner pops up when a user visits a website for the first time, requesting consent to the use of cookies and other trackers for a variety of purposes. Unlike prior work that has focused on evaluating the user interface (UI) design of cookie banners, this paper presents an in-depth analysis of \emph{what cookie banners say} to users to get their consent. We took an interdisciplinary approach to determining \textit{what cookie banners should say}. Following the legal requirements of the ePrivacy Directive (ePD) and the General Data Protection Regulation (GDPR), we manually annotated around 400 cookie banners presented on the most popular English-speaking websites visited by users residing in the EU. We focused on analyzing the purposes of cookie banners and how these purposes were expressed (e.g., any misleading or vague language, any use of jargon). We found that 89\% of cookie banners violated applicable laws. In particular, 61\% of banners violated the purpose specificity requirement by mentioning vague purposes, including ``\emph{user experience enhancement}’’. Further, 30\% of banners used positive framing, breaching the freely given and informed consent requirements. Based on these findings, we provide recommendations that regulators can find useful. We also describe future research directions.
\end{abstract}



\maketitle

\section{Introduction}
\input{1.introduction}\label{intro}

\section{Related work}
\input{2.related_work}\label{RelatedWork} 

\section{Methodology}
\input{3.Methodology}\label{METHOD} 

\section{Findings and discussion}
\input{4.Findings}\label{FINDINGS} 

\section{Conclusions and future work}
\input{5.Futurework}\label{FutureWork}

\begin{acks}
This work has been partially supported by the Luxembourg National Research Fund (FNR) -- IS/14717072 ``Deceptive Patterns Online (Decepticon)’’. We thank Nataliia Bielova, Claude Casteluccia, Rémy Coudert, and Michael Toth for their input and for designing and building the crawler used to create the dataset of banners.
\end{acks}

\bibliographystyle{ACM-Reference-Format}
\bibliography{bib/articles,bib/urls}

\section{Appendix}
\input{Appendix}\label{APPENDIX}

\end{document}

%% file: 1.introduction.tex
When users residing in the EU browse the web, they encounter a large number of banners prompting them to ``\emph{click accept to consent to the use of cookies as described in our cookie policy}’’. These banners appear because, according to the ePrivacy Directive (ePD)~\cite{ePD-09} and the General Data Protection Regulation (GDPR)~\cite{GDPR}, website operators, regardless of where they are based, must inform users located in the EU of the collection of their personal data. 
User consent is needed only when cookies and similar tracking technologies are used for \emph{unnecessary purposes}, such as advertising. 
Yet, website operators are required to be transparent and clearly explain the purpose of the use of cookies even if these cookies are necessary for the website to operate (i.e., \emph{necessary purposes}), such as authentication and security~\cite{EDPB-4-12}. Without knowing the specific purpose(s) of the use of cookies, users cannot decide whether to consent to the collection of their personal data. 

While prior studies have evaluated the user interface (UI) design of cookie banners to determine how design influences users’ consent decisions~\cite{bauer_are_2021,Gray-legalreq, Utz-etal-19-CCS,Nouwens-etal-20-CHI,Grassl2021dark}, little attention has been given to the textual elements of banners, which could likewise unlawfully steer users toward particular choices. There has been no in-depth analysis of what cookie banners say (and should say) and, hence, website owners can request user consent at their own discretion~\cite{Karegar-et-al, WP259}. They may employ technical jargon \cite{STRYCHARZ2021106750, Utz-etal-19-CCS}, vague and ambiguous language, and positive or negative framing. The resulting lack of transparency undermines users’ ability to understand why their data is collected and what risks are involved (Recital 39 GDPR, ~\cite{CJEU-Planet49-19,cnilvsgoogle2019,wp292016transparency}), hindering informed decisions and unlawfully nudging users toward giving their consent~\cite{Matt-etal-20-APF, NCC-Deceived-18,CNIL-Shap-19, CNIL-draft-rec-cookies-2020,EDPS-consumers,WPconsent2020}.

As part of ongoing work aimed at promoting transparency, lawfulness, and user-friendliness of cookie banner UIs, this paper addresses the following: \textit{What are the purposes of cookie banners, and how are these purposes expressed?} 
%
%
We combine expertise in data protection law, human-computer interaction (HCI), linguistics, and computer science to evaluate whether cookie banner text complies with the ePD and GDPR legal requirements concerning purposes and consent. To this end, we collected a corpus of about 1,300 cookie banners presented on the most popular English-speaking websites visited by users residing in the EU. We randomly selected and manually coded 407 of these banners, finding that 89\% of banners violated at least one legal requirement concerning processing purposes and consent. In particular, 20\% of banners violated the purpose availability requirement as they did not mention any processing purpose. More than 50\% of banners unlawfully mentioned the widely used but rather vague purpose: ``\emph{user experience enhancement}''. Further, 30\% used positive framing, breaching the freely given and informed consent requirements. Our findings suggest that many cookie banners use unlawful and questionable practices to obtain user consent. 

Our contributions are as follows:
\begin{enumerate}
\item We analyzed relevant legal sources and extracted six legal requirements explaining how cookie banner text should describe purposes of data collection and use;
\item We mapped legal requirements and their violations to observable linguistic features;
\item We empirically demonstrated that the wording used in cookie banners did not comply with ePD and GDPR;
\item We provided a set of recommendations that regulators and policymakers can find useful.
\end{enumerate}

\vspace{-2mm} 

%% file: 2.related_work.tex
Within the security and privacy domain, prior work has sought to assess the \emph{comprehensibility} of alert messages and warnings~\cite{W1, W2, W3, W4, W5, W6, W7, W8}, privacy policies~\cite{PP1, PP2}, contractual terms~\cite{murdoch2016payment,becker2017international}, browser disclosures~\cite{abu2020evaluating,abu2020designing}, and security and privacy advice on the web~\cite{redmiles2020comprehensive}. These studies have generally shown that privacy-related text is long and difficult to understand since it is overly complex and full of legalese \cite{Rossi2019when}, casting doubt on how informed users are when they make decisions with regard to the collection of their personal data. Moreover, the typical use of vague quantifiers (e.g., `certain’, `some’)~\cite{Reidenberg2016ambiguity} and modality markers (e.g., `may’, `might’) in privacy-related text makes it challenging for users to assess the data collection practices of organizations~\cite{Pollach2005typology}.
Further, scholars and regulators have shown that certain linguistic strategies may influence users’ online decisions by toying with users’ emotions \cite{Gray-etal-18-CHI}; e.g., shame \cite{Math-etal-19-HCI}, guilt \cite{Brig-Dark-18}, blame \cite{CNIL-Shap-19}, or fear \cite{Bongard2021manipulated}. 

With regard to the analysis of cookie banner text, Utz et al.~\cite{Utz-etal-19-CCS} showed that purposes were expressed in generic terms in almost half of the analyzed banners (e.g., ``\textit{to improve user experience}’’) and were unspecified in one out of six banners. In their empirical work, Hausner et al.~\cite{hausner2021dark} argued that positive framing (e.g., ``\textit{Yes, I am happy!}’’) could be used to nudge users toward giving their consent, whereas configuration options used to refuse or manage consent were expressed neutrally in banners. Similarly, Kampanos et al.~\cite{kampanos2021accept} showed that most banners presented ``affirmative’’ options that could nudge users toward consenting to tracking, whereas options like ``\textit{Read more}’’ and ``\textit{I do not accept}’’ were less prevalent. The remaining literature on how users’ consent to the use of cookies is requested exclusively focused on the UI design of cookie banners~\cite{bauer_are_2021,Gray-legalreq, Utz-etal-19-CCS,Nouwens-etal-20-CHI,Grassl2021dark}.

We build on prior work~\cite{Matt-etal-19-SP,Sant-etal-20-TechReg, Gray-legalreq} and employ user-centric transparency criteria \cite{wp292016transparency} to establish a benchmark that can be used to assess whether consent banner text is ePD- and GDPR-compliant. We use an inductive approach to investigating transparency issues (ambiguity, vagueness, technical jargon, misleading statements, and framing) through expert annotation of banner text.

\vspace{-1mm} 

%% file: 3.Methodology.tex


\noindent
\textbf{Data collection.} 
Using the Tranco list~\cite{LePo-etal-19-NDSS}, we created a dataset of about 1,300 cookie banners presented on the most visited English-speaking websites by users residing in the EU (in March 2020). We used the Polyglot library~\cite{polyglot} to detect the website language and then scrape English-speaking websites. The resulting set included both European and non-European domains. To \emph{scrape websites}, we used OpenWPM~\cite{englehardt2015openwpm}, a web privacy measurement framework based on Selenium~\cite{selenium}. It allowed full-page rendering before analysis and enabled taking screenshots of specific site elements. 

To \emph{detect cookie banners}, we followed three steps: segmentation, scoring, and tree traversal. First, we segmented webpages into small segments and built a segment tree~\cite{Math-etal-19-HCI} based on the segments’ HTML tag and text. Second, we assigned a score to each segment based on its inner text using a vocabulary set that we created by analyzing cookie banner content. 
We ranked tree leaf segments according to their scores. Third, we used the highest-scoring segments to traverse our segment tree. We performed bottom-up and top-down tree traversals. We captured HTML elements that contained cookie banners. To reduce false positives (i.e., websites with no banners), we used the segment scores to decide whether a cookie banner existed based on a threshold we set. 
We then manually filtered out any remaining false positives.

\noindent
\textbf{Legal requirements applicable to banner text.} \label{Elicitation} We analyzed various legal documents (ePD, GDPR, case law, regulatory decisions, and guidelines of non-binding sources like the European Data Protection Board (EDPB) and Data Protection Authorities (DPAs)) and extracted six legal requirements applicable to cookie banner text~\cite{Sant-etal-20-TechReg} (the requirements and their violations are described in Table~\ref{table:requirements} and further detailed in Table~\ref{METHOD}). The first two requirements (R1, R2) mandate that the \emph{purposes} of personal data processing should be described in an explicit and specific manner (Article 5(1)(b) GDPR)~\cite{Foua-etal-20-IWPE, WP203}. The other requirements (R3, R4, R5, R6) relate to the validity of \emph{consent}, which should be intelligible, expressed in clear and plain language, freely given, and informed (Articles 4(11), 7(2)(4) GDPR)~\cite{WPconsent2020}. Without explicit and specific purposes and without valid consent, websites may be found to infringe the GDPR’s principle of lawfulness (Article 6 (1)(a)), which would render any subsequent data processing unlawful and subject to heavy fines (Article 83 (5)(a) GDPR).
\vspace{-3mm} 
\input{TABLE}
\noindent

\textbf{Banner text coding.} The legal requirements we extracted (and their violations) were mapped to codes that we used to capture observable textual elements. Based on annotating our cookie banner set (described next), we identified five main codes: purpose of banners (see Table \ref{table:purposecodes}), framing, misleading language, vagueness, and technical jargon.

We -- a multidisciplinary team of five researchers with diverse expertise in data protection law, computer science, linguistics, and HCI -- iteratively coded a set of 150 banners (three iterations in total) using MAXQDA\footnote{MAXQDA: \url{https://www.maxqda.com/}}. We weekly met to develop our codebook until we reached good interrater agreement calculated for each pair of coders. Agreement ranged from 0.71 to 0.8 (Cohen's kappa coefficient) for all six pairs. We used the final codebook to annotate 407 banners that we randomly selected from the dataset we created. We analyzed the first layer of cookie banners without considering the second layer where the cookie policy can usually be found. Our choice was motivated by prior studies showing that many users simply disregard the second layer of consent requests~\cite{Utz-etal-19-CCS,McDonald_Cranor, Turow_Hennessy_Draper_2018} and make decisions exclusively based on the first layer. Additionally, transparency requirements mandate that the first layer of banners should give users a clear overview of data collection and processing~\cite{wp292016transparency} and that the second layer should be consistent with the first one \cite{wp292016transparency}.

\vspace{-1mm}

%% file: TABLE.tex
\begin{table}[htbp]
\small\addtolength{\tabcolsep}{-2pt}
\begin{tabular}{p{3.2cm}|p{4.8cm}} 
    \hline
    \textbf{Legal Requirement} & \textbf{Violation}\\
	\hline
	\textbf{R1 Purpose explicitness} & {}\\
	    R.1.1 Availability & Absence of purpose~\cite{wp292016transparency,CJEU-Planet49-19,HmbBfDI}\\
        R.1.2 Unambiguity & Ambiguous intent~\cite{WP203}\\
	    R.1.3 Shared common understanding & Inconsistent purposes~\cite{WP203}.\\
	\hline
	\textbf{R2 Purpose specificity} & Vague or general purposes \cite{WP203,wp292016transparency} \\
\hline
	\textbf{R3 Intelligible consent} & \\
	    R3.1 Non-technical terms & Presence of technical jargon \cite{wp292016transparency, SpanishDPA,cnilvsgoogle2019} \\
	    R3.2 Conciseness & Prolixity \cite{EDPBbyDefaultbyDesign,SpanishDPA,cnilvsgoogle2019} \\ \hline
	\textbf{R4 Consent with clear and plain language}     & {} \\
	    R4.1 Straightforward statements & Misleading expressions \cite{cnilvsgoogle2019,cnilvsamazon2020,SpanishDPA, wp292016transparency,garante-IT2020} \\
	    R4.2 Concreteness & Indefinite qualifiers \cite{WPconsent2020, wp292016transparency} \\ 
	\hline
	\textbf{R5 Freely given consent} & Pressure to provide consent \cite{WPconsent2020, wp292016transparency,CNIL-Shap-19,NCC-Deceived-18} \\ \hline
	\textbf{R6 Informed consent} & Absence of essential information about data processing~\cite{cnilvsgoogle2019, cnilvsamazon2020} \\ \hline
\end{tabular}
    \caption{A description of six legal requirements applicable to cookie banner text.}
    \label{table:requirements}
    \vspace{-8mm}
\end{table}

%% file: 4.Findings.tex
We present the findings of analyzing the text of 407 cookie banners. We also discuss the compliance of these banners with the six legal requirements we extracted. We found that 80\% of banners explained the purpose of data collection and processing. More than one-half of banners did not use misleading wording to explain purposes of processing, and about two-thirds of banners did not employ framing. Further, around 90\% of banners did not use vague language and technical jargon. However, 89\% of banners violated at least one out of six legal requirements considered in this study, as detailed below.

\noindent
\textbf{Absence of purposes.}
20\% of banners did not mention the purpose of data processing although several DPAs~\cite{ICO-Guid-19,FinishDPA,GreekDPA,BelgiumDPA} mandate transparent disclosure of purposes even for \emph{strictly necessary} cookies that do not require user consent. Hiding the reason for data processing violated the purpose availability (R1.1) and informed consent (R6) requirements.

\noindent
\textbf{Categories of purpose.} We identified eight different categories of purpose described in cookie banner text. We explain these categories in order of occurrence (from the most to the least recurring): user experience enhancement, analytics, advertising, custom content, service offering, essential functionalities, social media features, and profiling. The exact distribution of categories is shown in Table \ref{table:purposecodes}. We identified our categories by annotating banner text. Future work can map these categories to the ones created by several DPAs~\cite{SpanishDPA, ICO-Guid-19, WPconsent2020}.

\noindent
\textbf{Wording used to describe purposes.} We found a wide range of terms used to describe the purpose of data collection and processing, which often did not clearly match the eight categories we identified. For example, \emph{Advertising} was also referred to as marketing (content), targeted/tailored ads, ad(s) delivery/personalization/measurement, and promotional offers. \emph{Essential functionalities} were referred to as basic functions/functioning, operation of website, optimal website provision, and user preferences. Hence, the use of different terms to describe the same purpose did not comply with the common understanding (R1.3) requirement and raised the question about whether users could map different terms to the same concept.%

\noindent
\textbf{Most often mentioned purposes.} The \emph{user experience enhancement} purpose appeared in 61\% of banners that explained the purpose(s) of data processing. However, it was unclear how cookies improved the user experience of website visitors. It was also unclear whether certain cookies were necessary for the website to operate properly (e.g., adapting the presentation of website content to the user screen size~\cite{WP29-Cookies-13, EDPB-4-12}). DPAs and the EDPB~\cite{WP203, SpanishDPA} explain that such wording should not be used due to its vagueness and ambiguity. Thus, almost one-half of banners breached the specificity (R2), unambiguity (R1.2), and plain language (R4) requirements.

\emph{Analytics} was the second most mentioned purpose, appearing in 33\% of banners that described the purpose(s) of data processing. Since data could be collected and processed by first and/or third parties in aggregate or anonymously, the chosen wording violated the specificity (R2) requirement. We recall that third-party analytics entail the risk of cookie synchronization between different websites and, thus, that of profiling~\cite{CNIL-Carrefour}.

\emph{Profiling} was mentioned in 8\% of all banners that explained the purpose(s) of data processing, but it was rarely mentioned explicitly. More often, based on our interpretation of banner text, we recognized implicit mentions of the profiling purpose: ``\textit{[...] combine it with other information that you’ve provided to them or that they’ve collected from your use of their services}’’, violating the unambiguity (R1.2) and specificity of purposes (R2) requirements.

\noindent
\textbf{Multiple purposes.}
Some banners described a host of different processing purposes using a single sentence. For example, ``\emph{[…] to derive insights about the audiences who saw ads and content}’’ bundles up three purposes – analytics, advertising, and profiling – into one sentence. This violated the requirement of purpose specificity (R2) and possibly that of unambiguity (R1.2).

\begin{table*}[htbp]
\begin{tabular}{p{4.3cm}rp{11.3cm}}
\hline
Purpose category & Occur.  & Example\\
\hline
\textbf{User experience enhancement} & 61\% & \textit{[...] uses cookies to ensure you get the best experience on our website.}\\
\textbf{Analytics} & 33\%  & \textit{We use cookies to analyze our traffic.}\\ 
\textbf{Advertising} & 27\%  & \textit{Our site is using cookies for advertising purposes.}
\\
\textbf{Custom content} & 22\%  & \textit{This site uses cookies to help personalize content.}\\
\textbf{Service provision} & 15\% & \textit{This website uses cookies to provide its services.}\\ 
\textbf{Essential functionalities} & 14\%  & \textit{[...] uses cookies to ensure a comprehensive presentation and functionality of the website.} \\ 
\textbf{Social media features} & 11\%  & \textit{We use cookies to provide social media features.}\\
\textbf{Profiling} & 8\%  & \textit{This site use[s] profiling cookies to send you advertising based on your preferences.}\\
\hline
\end{tabular}
\caption{A description of eight data processing purposes we identified based on analyzing cookie banner text.} 
\label{table:purposecodes}
\vspace{-8mm}
\end{table*}

\noindent
\textbf{Misleading statements.} Misleading statements used to describe purposes were identified in 42\% of banners. They included descriptions that were vague, confusing, and ambiguous; were framed positively or negatively; instilled false beliefs; or concealed important information (i.e., deceiving the user~\cite{Boush2009deception}). Examples included the following: ``\textit{We use cookies that do not contain personal data about you in order to personalize content and ads}’’. Some statements were misleading due to framing, vagueness, or use of technical jargon.

\noindent
\textbf{Framing of purposes.} Positive framing was used in 30\% of banners describing purposes through the use of superlatives like `best’ and `most optimal’ (e.g., ``\textit{We use cookies to deliver the best possible web experience}’’). Positive framing was mainly used to describe the \emph{user experience enhancement} purpose, claiming that cookies optimized website performance or improved user experience. Highlighting the positive aspects of consenting to cookie processing provided a partial view, making users pay less attention to other aspects that could be deemed negative (e.g., targeted advertising)~\cite{Boush2009deception} but key to making informed decisions. Thus, positive framing violated the freely given (R5) and informed consent (R6) requirements.

Negative framing was only used in 2\% of banners, mainly to warn users of the loss of functionalities if users did not consent to the use of cookies; e.g., ``\textit{If you’re not happy with this, we won’t set these cookies but some nice features of the site may be unavailable}’’. When choices are framed negatively, they may put pressure on users by exploiting loss aversion \cite{Acquisti2017nudges} and nudge them toward consenting~\cite{CNIL-Shap-19}, especially when it is unclear which functionalities will be lost. Our previous study \cite{Bongard2021manipulated} showed that people may develop wrong mental models of the consequences of (not) consenting to data collection and processing. Therefore, both positive and negative framing may nudge users toward complying with the service provider’s wishes \cite{NCC-Deceived-18}, violating the freely given consent (R5) requirement.

\textbf{Necessary vs. unnecessary cookies.} Additionally, we found that cookie banner text did not explain the difference between necessary and unnecessary cookies. Therefore, non-expert users may be misled to believe that all cookies are necessary for websites to operate properly~\cite{Solove2013privacy,Ben2011failure, STRYCHARZ2021106750}. For example, ``\textit{This website or its third-party tools use cookies, which are necessary to its functioning and required to achieve the purposes illustrated in the cookie policy}'' is a misleading statement because it claims that all cookies are key to proper website functionality, making users' consent uninformed and violating the specificity (R2), unambiguity (R1.2), clear and plain language (R4), and informed consent (R6) requirements.

\noindent
\textbf{Vagueness.} 11\% of banners used vague terms to describe purposes: ``\textit{We may share information about your use of the site with third parties; we may use cookies}’’. Vagueness is \textit{misleading} when it leaves individuals uncertain about the intended meaning of an expression~\cite{Rossi2019when,Pollach2005typology}, in particular whether cookies are used and whether data is shared with other parties. The \emph{user experience enhancement} and \emph{service provision} purposes -- mentioned cumulatively in more than one-half of banners -- could also be deemed misleading if proven untrue, especially that vague language does not explain how accepting cookies is beneficial to users~\cite{WP203, Matt-etal-20-APF, cnilvsgoogle2019, SpanishDPA}.

\noindent
\textbf{Technical jargon and prolixity.} The description of purposes in 9\% of banners contained technical jargon, including ``\textit{derive insights about the audiences}’’, ``\textit{retargeting cookies}’’, and ``\textit{Google Analytics}’’, breaching the requirements of intelligibility (R3) and straightforward statements (R4.1). While examples or explanations can clarify technical terms, they can still breach the conciseness (R3.2) requirement given the limited size of cookie banners. Empirical studies should be conducted to assess the comprehensibility of technical terms by different audiences, although some prior studies~\cite{Utz-etal-19-CCS,STRYCHARZ2021106750,Bosc-etal-16-PETs, Santos2017DetectingAE} have shown that the use of technical jargon could leave users in a vulnerable state.

\noindent
\textbf{Recommendations.} Based on our findings, we provide recommendations that regulators and policymakers can find useful.

\textit{Standardization of purposes.} Given that people may have different interpretations of the same text, we argue that most consensual purposes (e.g., advertising, statistical analysis, social media features, personalization) and their labels should be standardized, following Privacy by Design~\cite{richter2018entrepreneurial}. The violations we identified in this work are rooted in the fact that websites can describe purposes at their own discretion. The EDPB, DPAs, and standard committees should standardize purpose categories to minimize legal uncertainty and simplify data processing operations. 


\textit{Requirements for describing purposes.} The legal requirements that currently exist are generic and difficult to operationalize. Therefore, there is a need for a set of requirements that can be used to help define purposes of data processing. We also argue that creating a blacklist of illegal purposes could improve the current situation. A simple nomenclature like `\emph{necessary}’ and `\emph{unnecessary}’ cookies could improve users’ comprehensibility of cookie banner text and help users make informed decisions, provided that details about purposes of data processing can be found in cookie policies. Yet, from a user’s point of view, we believe the best solution would be managing cookie consent options pre-emptively at the browser level or through an automated browser extension.

\textit{Language tensions.} Best practices and examples of how to clearly refer to data privacy concepts while remaining concise should also be made available to website providers, since attempts to comply with the explicitness, specificity, and plain language legal requirements could also lead to prolixity.

\vspace{-2mm}

%% file: 5.Futurework.tex
In this work, we analyzed the text describing the data processing purposes of 407 cookie banners. We found that 89\% of banners violated at least one legal requirement. 20\% of banners did not mention any purpose although purpose disclosure is legally mandated. Notably, 67\% of banners violated the specificity requirement, and 61\% unlawfully provided a vague purpose: ``\emph{user experience enhancement}’’. 31\% of banners used framing, breaching the freely given consent requirement. Other identified issues included misleading statements, technical jargon, and vagueness.

We argue that the identified violations do not allow users to be aware of the scope, consequences, and risks (Recital 39 GDPR) of consenting to storing cookies on their devices, especially the privacy-invasive ones and, as a result, breaching the principle of transparency that governs personal data processing (Article 5(1)(a) GDPR). It is not only difficult for lay users to understand cookie banner text, but also experts may find it challenging to parse banner text and map it to relevant legal requirements. This suggests that, besides necessary standardization, a purpose-based consent may neither be user-friendly nor feasible and, hence, we argue that unnecessary cookies should be rejected by default.
%

We will build on this work and conduct user studies to empirically evaluate the comprehensibility of different textual elements of banners. We will investigate the influence of positive and negative framing on users’ choices. We will also seek to create a taxonomy of commonly understandable purposes to facilitate comprehensibility, comparability, and compliance checking. It would also be useful to use natural language processing (NLP) to automate, for example, the detection of misleading text as well as use sentiment analysis to identify positive and negative framing.

%% file: Appendix.tex
The table on the next two pages describes the six legal requirements we extracted from legal sources, what constituted a violation of these requirements, and how we mapped these requirements (and their violations) to the codebook we developed based on our cookie banner text annotation.

\begin{table*}[ht]
\small\addtolength{\tabcolsep}{-1pt}
\begin{tabular}{p{3cm} p{3.2cm} p{3.2cm} p{3cm} p{4cm}}
    \hline
    \textbf{Requirement} & {\textbf{Requirement definition}} & \textbf{Violation} & \textbf{Code (from codebook)} & \textbf{Code definition}\\
	\hline
	\textbf{R1 Purpose explictiness} & {} & {} & {} & {} \\
	R.1.1 Availability & Purposes should be clearly expressed, revealed, or explained, especially on the first layer of consent banners \cite{HmbBfDI}. & Absence of purpose. This requirement stems from the transparency principle (Article 5(1)(a), Recital 39 GDPR); data controllers need to inform users of data processing purposes (Article 13 (1)(c) and Recitals 58, 60 GDPR). & Code: purpose & We applied code when banners did not explain the purpose(s) of data processing.\\
    R.1.2 Unambiguity & Purposes should be unambiguous. & Ambiguous intent. A violation occurs when purposes are defined ambiguously, and there is doubt about their meaning or intent \cite{WP203}. & Codes: vagueness; prolixity; positive framing; negative framing \cite{WP203} & We applied code when banners used ambiguous wording.\\
	R.1.3 Shared common understanding & Purposes should be comprehensible, regardless of users’ cultural or linguistic backgrounds or other special needs involved \cite{WP203}. & Purposes are not comprehensible. & Code: purpose & We applied code when different terms were used to describe the same purpose.\\
	\hline
	\textbf{R2 Purpose specificity} & Purposes should be precisely identified, clearly defined, and detailed enough to determine what kind of processing is included or excluded within the specified purpose \cite{WP203, wp292016transparency}. & Violations occur when a purpose is too vague or generic; for instance, ``improve users' experience’’; ``develop new services and products’’; ``offer personalized services'' \cite{WP203, wp292016transparency}. & Code: purpose; sub-codes: user experience enhancement; analytics; advertising; custom content; service provision; essential functionalities; social media features; profiling & We applied sub-codes based on specific data processing purposes banners described.\\
    \\ \hline
	\textbf{R3 Intelligible consent} & {} & {} & {} \\ 
	R3.1 Non-technical terms & Consent should not contain overly legalistic or technical language. & Use of technical jargon. & Code: technical jargon \cite{wp292016transparency, SpanishDPA, cnilvsgoogle2019}
	& We applied code when banners used terms that non-expert users are usually unfamiliar with (e.g., JavaScript, trackers, tracking systems, clients, servers).\\
	R3.2 Conciseness & The first layer of consent requests should be brief but contain sufficient information~\cite{EDPBbyDefaultbyDesign}. & A violation occurs when unnecessary details are mentioned, distracting users or causing information overload~\cite{SpanishDPA,cnilvsgoogle2019}. & Code: prolixity & We applied code when unnecessary details were mentioned in banners. \\ \hline
	\textbf{R4 Consent with clear and plain language} & {} & {} & {} \\
	R4.1 Straightforward statements & Consent requests should describe purposes clearly \cite{WPconsent2020}. & A violation occurs when misleading expressions are used: ``We use cookies to personalize content and create a better user experience’’ \cite{SpanishDPA, wp292016transparency,garante-IT2020, cnilvsgoogle2019,cnilvsamazon2020}. & Code: vagueness & We applied code when indefinite qualifiers were used in banners: can, may, might, someone, certain information, data, other parties, our vendors, some, any, often, possible, etc. Examples include ``We may place cookies …''; ``We may use information for …’’ \cite{WPconsent2020, wp292016transparency}.\\
	R4.2 Concreteness & Consent requests should use accurate and definitive statements. & A violation occurs when deceptive language or indefinite qualifiers are used \cite{WPconsent2020, wp292016transparency}. & Code: vagueness & We applied code when banners did not use clear language and, instead, used indefinite qualifiers, incomplete or ambiguous statements, double negatives, or deceptive practices by hiding information from users (e.g., whether or not certain cookies are necessary for the website to operate properly).\\
	\end{tabular}
\end{table*}

\begin{table*}[ht]
\small\addtolength{\tabcolsep}{-1pt}
\begin{tabular}{p{3cm} p{3.2cm} p{3.2cm} p{3cm} p{4cm}}
    \hline
    \textbf{Requirement} & {\textbf{Requirement definition}} & \textbf{Violation} & \textbf{Code (from codebook)} & \textbf{Code definition}\\
	\hline
	\textbf{R5 Freely given consent} & A request for consent should imply a voluntary choice to accept or decline the processing of personal data (Articles 4 (11), 7(4) GDPR). & Any sort of pressure that nudges users toward consenting \cite{WP259} (e.g., the use of positive or negative framing~\cite{CNIL-Shap-19,NCC-Deceived-18}). & Code: positive framing; sub-codes: assumed happiness; safety and privacy arguments; compliance and authority arguments; playful arguments; superlatives and better experiences & We applied codes when positive or negative framing was used.\\
	&&& Code: negative framing; sub-codes: worse user experience; loss of functionalities &\\
	\hline

	\textbf{R6 Informed consent} & When trackers are used and stored on users’ devices, users must be informed and aware of these trackers. & Absence of essential information about data processing. & Code: data type & We applied code when banners mentioned the types of data collected – IP address; geolocation data. \\ \hline
    \end{tabular}
    \caption{Mapping the six legal requirements we extracted from different legal sources (and their violations) to our codebook, which we developed based on annotating the text of 407 cookie banners.}
    \label{table:method}
\end{table*}